\documentclass[a4paper, oneside, twocolumn, notitlepage, 10pt]{extarticle_ecoc}
\usepackage{ecoc}
\usepackage{comment}
\usepackage{color}
\usepackage{subfigure}
\usepackage{enumerate}
\usepackage{enumitem}
\usepackage{stfloats}
\usepackage{psfrag}
\usepackage{setspace}
\usepackage{balance}
\usepackage{enumitem}
\usepackage{tikz}
\usepackage{pgfplots,amsfonts}
\usepackage{pgfplotstable}
\usepackage{amsmath,amssymb}
\usepackage{xcolor}
\usepackage{graphicx}
\usepackage{tikz}
\usepackage{pgfplots}
\usepackage{float}
\usepackage{balance}
\usepackage{setspace}
\usepackage{xspace}
\usepackage[normalem]{ulem}
\usepackage{pifont}
\usepackage{dsfont,empheq}
\usepackage{multicol,multirow}

\usepackage{booktabs, cellspace, hhline}

\usetikzlibrary{shapes}
\usetikzlibrary{spy}
\usetikzlibrary{fit}
\usetikzlibrary{shapes.multipart}
\usetikzlibrary{positioning}
\pgfplotsset{compat=1.14}
\usepackage{wrapfig}
\usepackage[acronym,nomain]{glossaries}

\newacronym{CUT}{CUT}{channel under test}

\newacronym{KK}{KK}{Kramers-Kronig}
\newacronym{CSPR}{CSPR}{carrier-to-signal power ratio}
\newacronym{KKRX}{KKRx}{Kramers-Kronig Receiver}
\newacronym{SSBI}{SSBI}{signal-signal beat interference}

\newacronym{GS}{GS}{geometric shaping}
\newacronym{PS}{PS}{probabilistic shaping}
\newacronym{DSP}{DSP}{digital signal processing}
\newacronym{MIMO}{MIMO}{multiple-input multiple-output}
\newacronym{TDE}{TDE}{time domain equalizer}
\newacronym{FDE}{FDE}{frequency domain equalizer}
\newacronym{LMS}{LMS}{least means square}
\newacronym{DDLMS}{DD-LMS}{decision directed least means square}
\newacronym{FFE}{FFE}{feed-forward equalizer}
\newacronym{FBE}{FBE}{feedback equalizer}
\newacronym{BPS}{BPS}{blind phase search}

\newacronym{SMF}{SMF}{single-mode fiber}
\newacronym[plural=SSMFs]{SSMF}{SSMF}{standard single-mode fiber}
\newacronym[plural=FMFs]{FMF}{FMF}{few-mode fiber}
\newacronym{FMF12}{FMF12}{12 mode FMF}
\newacronym{MMF}{MMF}{multi-mode fiber}
\newacronym{SI}{SI}{step index}
\newacronym{GI}{GI}{graded index}
\newacronym{DCF}{DCF}{dispersion compensated fiber}

\newacronym{SDM}{SDM}{space division multiplexing}
\newacronym{MDM}{MDM}{mode division multiplexed}
\newacronym{WDM}{WDM}{wavelength division multiplexing}
\newacronym{DWDM}{DWDM}{dense wavelength division multiplexing}
\newacronym{LP}{LP}{linear polarized}
\newacronym[plural=MMUXs,firstplural=mode multiplexers]{MMUX}{MMUX}{mode multiplexer}
\newacronym{PL}{PL}{photonic lantern}
\newacronym{3DWG}{3DWG}{3D-waveguide}
\newacronym{MDL}{MDL}{mode dependent loss}
\newacronym{DGD}{DGD}{differential group delay}
\newacronym{DMGD}{DMGD}{differential mode group delay}
\newacronym{QSM}{QSM}{quasi-single-mode}
\newacronym{GIMMF}{GI-MMF}{graded-index multi-mode fiber}

\newacronym{SSB}{SSB}{single side band}
\newacronym{QPSK}{QPSK}{quadrature phase shift keying}
\newacronym{QAM}{QAM}{quadrature amplitude modulation}
\newacronym{RRC}{RRC}{root-raised-cosine}
\newacronym{4D-64PRS}{4D-64PRS}{
four-dimensional 64-ary polarization-ring-switching}
\newacronym{DP}{DP}{dual-polarization}
\newacronym[\glslongpluralkey=states-of-polarization]{SOP}{SOP}{state-of-polarization}
\newacronym{PM}{PM}{polarization-multiplexed}

\newacronym{ECL}{ECL}{external cavity laser}
\newacronym{CW}{CW}{continuous wave}
\newacronym[plural=DFBs]{DFB}{DFB}{distributed feedback laser}

\newacronym[plural=DACs]{DAC}{DAC}{digital-to-analog converter}
\newacronym{ADC}{ADC}{analog-to-digital converter}
\newacronym{PRBS}{PRBS}{pseudo-random bit sequence}
\newacronym{LO}{LO}{local oscillator}
\newacronym{EDFA}{EDFA}{erbium-doped fiber amplifier}
\newacronym{MZM}{MZM}{Mach-Zehnder modulator}
\newacronym{DP-MZM}{DP-MZM}{dual-polarization Mach-Zehnder modulator}
\newacronym{ChUT}{ChUT}{channel under test}
\newacronym{WSS}{WSS}{wavelength selective switch}
\newacronym[plural=VOAs]{VOA}{VOA}{variable optical attenuator}
\newacronym[plural=PDCRXs]{PDCRX}{PDCRX}{polarization diverse coherent receiver}
\newacronym{DSO}{DSO}{digital storage oscilloscope}
\newacronym{ASE}{ASE}{amplified spontaneous emission}
\newacronym{PBS}{PBS}{polarization beam splitter}
\newacronym{PD}{PD}{photodiode}
\newacronym{AOM}{AOM}{acousto-optic modulator}
\newacronym{BPD}{BPD}{balanced photo-diode}
\newacronym{OMFT}{OMFT}{optical multi-format transmitter}
\newacronym{DPIQ}{DP-IQM}{dual-polarization IQ-modulator}
\newacronym{ABC}{ABC}{automatic bias controller}
\newacronym{OTF}{OTF}{optical tunable filter}
\newacronym{LSPS}{LSPS}{loop-synchronous polarization scrambler}
\newacronym{OSA}{OSA}{optical spectrum analyzer}

\newacronym{OSNR}{OSNR}{optical signal to noise ratio}
\newacronym{BER}{BER}{bit error rate}
\newacronym{IL}{IL}{insertion loss}

\newacronym{SDFEC}{SD-FEC}{soft-decision forward error correction}
\newacronym{HDFEC}{HD-FEC}{hard-decision forward error correction}
\newacronym{FEC}{FEC}{forward error correction}
\newacronym{LDPC}{LDPC}{low-density parity-check code}

\newacronym{AIR}{AIR}{achievable information rate}
\newacronym{AR}{AR}{achievable rates}
\newacronym{MI}{MI}{mutual information}
\newacronym{GMI}{GMI}{generalized mutual information}
\newacronym{BICM}{BICM}{bit-interleaved coded modulation}
\newacronym{GN}{GN}{Gaussian noise}

\newacronym{OVNA}{OVNA}{optical vector network analyzer}
\newacronym{NIR}{NIR}{near infrared}
\newacronym{CD}{CD}{chromatic dispersion}
\newacronym{OTDR}{OTDR}{optical time domain reflectometry}
\newacronym{OFDR}{OFDR}{optical frequency domain reflectometry}
\newacronym{GPU}{GPU}{graphics processing unit}
\newacronym{SVD}{SVD}{singular value decomposition}
\newacronym{WGN}{WGN}{white Gaussian noise}
\newacronym{AWGN}{AWGN}{additive white Gaussian noise}
\newacronym{PDL}{PDL}{polarization dependent loss}
\newacronym{SPS}{sps}{samples-per-symbol}
\newacronym{SE}{SE}{spectral efficiency}

\usepackage{arydshln}

\definecolor{myDarkGreen}{rgb}{0.00000,0.58824,0.00000}%

\definecolor{yellow}{RGB}{250, 199, 100} 
\definecolor{blue}{rgb}{0.38, 0.51, 0.71} 
\definecolor{darkblue}{RGB}{17, 42, 60} 
\definecolor{red}{RGB}{175, 49, 39} 

\definecolor{orange}{RGB}{217, 156, 55} 
\definecolor{green}{RGB}{144, 169, 84} 
\definecolor{palegreen}{RGB}{197, 184, 104} 

\definecolor{yellow}{RGB}{250, 199, 100} 
\definecolor{brokenwhite}{RGB}{218, 192, 166} 
\definecolor{brokengrey}{rgb}{0.77, 0.76, 0.82} 

\usetikzlibrary{patterns}

\begin{document}
\selectlanguage{english}    


\title{On the Performance of Multidimensional Constellation Shaping for Linear and Nonlinear Optical Fiber Channel\footnote{This paper is a preprint of a paper submitted to and accepted for publication in ECOC 2023 and is subject to Institution of Engineering and Technology Copyright. The copy of record will be available at IET Digital Library}}


\author{
     Bin Chen\textsuperscript{(1),}*, Zhiwei Liang\textsuperscript{(1)}, Shen Li\textsuperscript{(2)},  Yi Lei\textsuperscript{(1)},  Gabriele Liga\textsuperscript{(3)} 
     and Alex Alvarado\textsuperscript{(3)}
}

\maketitle                  


\begin{strip}
 \begin{author_descr}

   \textsuperscript{(1)}  School of Computer Science and Information Engineering, Hefei University of Technology, China
   
   * \textcolor{blue}{\uline{bin.chen@hfut.edu.cn}}

 \textsuperscript{(2)} Department of Electrical Engineering, Chalmers University of Technology, Gothenburg, Sweden

  \textsuperscript{(3)} Department of Electrical Engineering, Eindhoven University of Technology, The Netherlands

 \end{author_descr}
\end{strip}

\setstretch{1.062}

\begin{strip}
  \begin{ecoc_abstract}
  {Multidimensional constellation shaping of up to 32 dimensions with different spectral efficiencies are compared through AWGN and fiber-optic simulations. The results show that no constellation is universal and the balance of required and effective SNRs should be jointly considered for the specific optical transmission scenario.} 
 \copyright 2023 The Author(s) 
  \end{ecoc_abstract}
\end{strip}

\section{Introduction}
The application of multidimensional (MD) modulation in optical
fiber communications has been intensively investigated since the fiber-optic channel offers different physical dimensions of light waves (quadratures, polarizations,  wavelengths, modes and  cores) simultaneously \cite{AgrellJLT2009,Karlsson:09}. 
Compared to the typical paradigm of transmitting independent data on all of these dimensions, improved performance can be obtained by optimizing the distribution of the constellation points or encoding data jointly on several dimensions, i.e., by \textit{multidimensional constellation shaping} \cite{ForneyJSAC1984,Magnus2016}.

To harvest performance gains in optical transmission systems, the optimization of novel MD modulation formats has attracted considerable attention for reducing nonlinear interference (NLI) distortions in 4D \cite{Chagnon:13,Kojima2017JLT,BinChenJLT2019,SebastiaanJLT2023,GabrieleOFC2022,EricJLT2022,Ling2022}, 8D \cite{Shiner:14,El-RahmanJLT2018,Mirani2021,BinChenPTL2019}, and higher dimensions (up to 32D \cite{Li2023}). The key idea relies on the fact that MD shaping can reduce the NLI generated by  the variations of the transmitted signal energy in addition to the usual (linear) shaping gain, and it could potentially be higher than the ultimate shaping gain (1.53~dB) in additive white Gaussian noise (AWGN) channel \cite{DarISIT2014,OmriJLT2016}. {However, the problem of designing optimal MD constellations with a finite number of dimensions  for arbitrary fiber-optic channels is still open.}
  
In this paper, we analyze the performance of various MD modulation formats with geometric shaping (GS) for  both the AWGN channel and  the optical fiber channel. By evaluating the dependency of the effective signal-to-noise ratio (SNR)  on the MD modulation format in multi-span and single-span transmission systems with the help of a recently proposed 4D NLI model  \cite{2020Extending,liang2023}, {we show that the MD formats should be evaluated by combing the increasing shaping gain in AWGN channel and the enhanced  effective SNR in optical fiber channels.
}

\vspace{-0.7em}
\section{Performance Metrics: MI, GMI and SNRs}
As the most popular information-theoretical performance metrics,
mutual information (MI) and generalized MI (GMI) 
can quantify the maximum number of information bits per transmit symbol in symbol-wise and bit-wise coded modulation systems, respectively \cite{AlvaradoJLT2018}. ~For arbitrary $M$-ary $N$-dimensional modulation formats with equal probabilities that can carry at most $m=\frac{4}{N}\log_2 M$ bit/4D-symbol as spectral efficiency (SE), 
the  maximum number of information bits per transmit bit can be obtained by normalizing the MI or GMI as $\text{NMI}=\text{MI}/m$ or $\text{NGMI}=\text{GMI}/m$, where $0\leq \text{NGMI}\leq \text{NMI} \leq 1$. The values of NMI and NGMI indicate the largest ideal code rate that is able to produce error-free post-FEC results.

MD constellation shaping aims at   reducing the required SNR over the AWGN channel but in practice often leads to a smaller received (effective)  SNR  due to hardware distortions and fiber nonlinearities. Thus, true performance improvement of MD  formats in optical channel is achieved if and only if the reduction in the required SNR is larger than the loss in terms of the effective SNR. The required SNR  is defined as the minimum SNR to achieve a target value of a performance metrics (here NMI and NGMI).

The analysis in this paper is based on a set of modulation formats, which combines the MI or GMI-optimized
2D and 4D formats in our online database\cite{Database-Github} and MD Voronoi constellations (VCs) in \cite{Li2022, Li2023}.
{In the choice of MD modulation format, there is an inherent threefold trade-off among the SE, the noise tolerance, and the complexity of the format \cite{ChenJLT2023}.}
Thus, the performance of a  modulation format  is heavily dependent on which channel and performance metric is considered.

\vspace{-0.7em}
\section{Results: Shaping Gap to AWGN Capacity}
In this section, we  perform extensive MI and GMI evaluations for the considered  formats \footnote{The numerical results in this paper are based on a set of modulation formats which combines the MI or GMI-optimal 2D and 4D  formats in  \cite{Database-Github} and  MD Voronoi constellations (VCs) in \cite{Li2022, Li2023}.}
 in the AWGN channel.  For the AWGN channel, the performance of an ideally shaped QAM system matches that of an ideal system using Gaussian modulation, which represents the case of a Gaussian input distribution (equivalent to the case of $N\rightarrow\infty$) \cite{DarISIT2014}.

\begin{figure}[!tb]
    \centering
    {\includegraphics{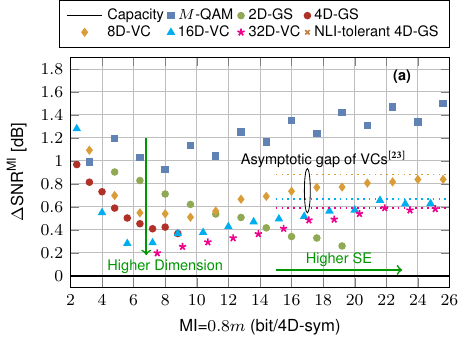}}
    {\includegraphics{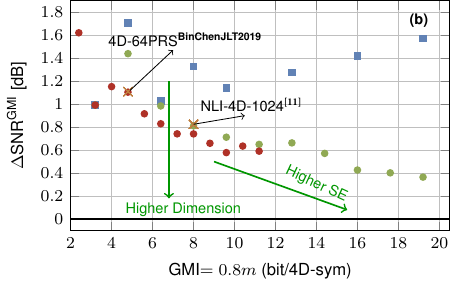}}
        \caption{Comparison of the gap to AWGN capacity for different MD  formats at the target rates: (a) NMI=0.8  or  (b) NGMI=0.8.}
    \vspace{-0.3em}
    \label{fig:CapacityGap}
\end{figure}

To quantify the gains offered by different modulations, we plot the ``SNR gap to AWGN capacity" ($\Delta\text{SNR}$) in Fig.~\ref{fig:CapacityGap}. The value of $\Delta\text{SNR}$ in dB measures how far a given symbol-wise or  bit-wise coded modulation scheme is operating from the Gaussian modulation, which are defined as
\begin{align}\Delta\text{SNR}^R_{\text{req}}=\text{SNR}_{\text{req}}^R-\text{SNR}_{\text{req}}^{C},
\end{align}
where the $R$ is MI or GMI and $C$ is Shannon capacity. Assuming a coded modulation system with an ideal soft-decision FEC with $R_c=0.8$ code rate into account, the required SNR is obtained for $R=0.8m$ and $C=0.8m$, where the $m$ is the number of bits of the constellation. 

Fig.~\ref{fig:CapacityGap} (a) and (b) collectively indicate that (i) MD geometric shaping can provide significant gains compared to uniform QAM, (ii) by increasing the dimension or SE, the SNR gap of constellations can be brought closer to the Gaussian capacity as shown in Table.~\uppercase\expandafter{\romannumeral3} \cite{ForneyJSAC1984}, (iii) the 4D-GS constellation (red  circle \tikz{\draw plot[mark=*, mark size=2,mark options={color=red, thick}] (,);}) exhibits a larger shaping gain than the 2D-GS constellation (green circle \tikz{\draw plot[mark=*, mark size=2,mark options={color=green, thick}] (,);)}), (iv) for the SE of a 4D constellation with odd bits, a single 2D constellation does not exist, i.e., 4D formats achieve a finer granularity than 2D formats. 

For MD-VCs (8D-VC \tikz{\draw plot[mark=diamond*, mark size=2.5,mark options={color=orange, thick}] (,);}, 16D-VC \tikz{\draw plot[mark=triangle*, mark size=2,mark options={color=cyan, thick}] (,);} and 32D-VC \tikz{\draw plot[mark=star, mark size=2.5,mark options={color=magenta, thick}] (,);}), the $\Delta\text{SNR}^{\text{MI}}_{\text{req}}$ decreases as the dimension increases, as shown in Fig.~\ref{fig:CapacityGap} (a). 
As the SE increases, the $\Delta\text{SNR}^{\text{MI}}_{\text{req}}$ converges to the asymptotic gap $1.53 - \gamma_{\text{s}}$ [dB] (dotted lines), where $\gamma_{\text{s}}$ is the asymptotic shaping gains of the shaping lattices listed in \cite[Table~I]{Li2022}. Compared to 2D-GS and 4D-GS formats, 16D- and 32D-VCs realize lower $\Delta\text{SNR}^{\text{MI}}_{\text{req}}$ between 4 and 14 bit/4D-sym, and provide a good trade-off at high SEs due to their low-complexity encoding and decoding algorithms \cite{Li2022}. 
However, for MD-VCs finding efficient labelings which lead to a better performance than QAM at NGMI$=0.8$ is challenging, thus the $\Delta\text{SNR}^{\text{GMI}}_{\text{req}}$ results for MD-VCs is not shown in Fig.~\ref{fig:CapacityGap} (b).

{Lastly, two nonlinear-tolerant 4D formats (4D-64PRS \cite{BinChenJLT2019} and NLI-4D-1024 \cite{Ling2022}) are highlighted as cross marks \tikz{\draw plot[mark=x, mark size=2.5,mark options={color=brown, thick}](, );}  for GMI of 4.8~bit/4D-sym  and 8~bit/4D-sym in Fig.~\ref{fig:CapacityGap} (b).  These two constellations lead to a small but negligible  loss compared to the AWGN-optimized 4D-GS formats. However, more gains are expected in the nonlinear optical fiber channel (see Fig.~\ref{fig:NLI_results} in next section).} 
 
\vspace{-0.7em}
\section{Results: Shaping Gains in Nonlinear Optical Fiber Channel}
The optical system under consideration is a 11-channel WDM  system with  96~GBaud symbol rate  and 100~GHz channel spacing. For getting NMI or  NGMI approximately at 0.8$m$, two transmission cases with standard single-mode fiber, 60$\times$80~km multi-span system and a 205~km single-span system, are considered. The fiber parameters are a loss coefficient of 0.2~dB/km, a  chromatic dispersion of 17~ps/nm/km, and a nonlinear coefficient of 1.3~/W/km. Fiber losses are compensated after each span by erbium-doped fiber amplifiers, with a noise figure of 5~dB.

\begin{figure}[!tb]
  \vspace{-1em}
    \centering
    \input{tex/1}
    \caption{The effective SNR performance comparison  over two different transmission cases via 4D NLI model  \cite{liang2023}. The considered constellations are shown in Table.~\ref{tab:NLI_SNR}.}
    \label{fig:NLI_results}
\end{figure}

\begin{table}[!tb]
  \vspace{0.5em}
		\centering
		\caption{The $\text{SNR}_{\text{req}}^R$ and $\text{SNR}_{\text{eff}}$ of the constellations in Fig.~\ref{fig:NLI_results}.}
   \vspace{-0.1em}
		\scalebox{0.715}{
		\begin{tabular}{c|l|c|c|c}
\hline
\hline
\hspace{-0.3em}SE ($m$)& \hspace{3em} Constellation & $\text{SNR}_{\text{req}}^{\text{MI}}$ & $\text{SNR}_{\text{req}}^{\text{GMI}}$ & $\text{SNR}_\text{eff}$\\
\hline
\hline
\multirow{7}{*}{6}
& \textcircled{\tiny{1}} 8QAM (\tikz{\draw [solid, color = blue] (1, 1) -- (1.5, 1); \draw plot[mark=square*, mark size=1.5,mark options={color=blue, thick}] (1.25, 1);})& 7.502 & 8.117 &  10.995\\

\cline{2-5}
& \textcircled{\tiny{2}} hepta2-8 (\tikz{\draw [solid, color = green] (1, 1) -- (1.5, 1); \draw plot[mark=*, mark size=1.5,mark options={color=green, thick}] (1.25, 1);})& 7.215 & 7.830 & 11.090 \\

\cline{2-5}
&\textcircled{\tiny{3}} C4-64 (\tikz{\draw [solid, color = red] (1, 1) -- (1.5, 1); \draw plot[mark=*, mark size=1.5,mark options={color=red, thick}] (1.25, 1);})& 6.901 & 8.835 & 11.078\\

\cline{2-5}
& \textcircled{\tiny{4}} DSQ2-8 (\tikz{\draw [dashed, color = green] (1, 1) -- (1.5, 1); \draw plot[mark=*, mark size=1.5,mark options={color=green, thick}] (1.25, 1);}) & 7.332 & 7.752 & 11.005 \\

\cline{2-5}
&\textcircled{\tiny{5}} GS-AWGN-4D-64 (\tikz{\draw [dashed, color = red] (1, 1) -- (1.5, 1); \draw plot[mark=*, mark size=1.5,mark options={color=red, thick}] (1.25, 1);}) & 7.242 & 7.417 & 10.962 \\


\cline{2-5}
&\textcircled{\tiny{6}} 4D-64PRS (\tikz{\draw [dashed, color = brown] (1, 1) -- (1.5, 1); \draw plot[mark=x, mark size=2,mark options={color=brown, thick}] (1.25, 1);}) & 7.257 & 7.421 & 11.166 \\

\cline{2-5}
& Gaussian distribution & 
\multicolumn{2}{c|}{$\text{SNR}_{\text{req}}^{C}=6.312$}  & 10.705\\

\hline
\multirow{6}{*}{10}
&\textcircled{\tiny{7}} 32QAM (\tikz{\draw [solid, color = blue] (1, 1) -- (1.5, 1); \draw plot[mark=square*, mark size=1.5,mark options={color=blue, thick}] (1.25, 1);})& 12.686 & 13.091 & 13.119\\


\cline{2-5}
&\textcircled{\tiny{8}} GS-AWGN-2D-32 (\tikz{\draw [dashed, color = green] (1, 1) -- (1.5, 1); \draw plot[mark=*, mark size=1.5,mark options={color=green, thick}] (1.25, 1);}) & 12.472 & 12.572 & 12.841\\


\cline{2-5}
&\textcircled{\tiny{9}} C4-1024 (\tikz{\draw [solid, color = red] (1, 1) -- (1.5, 1); \draw plot[mark=*, mark size=1.5,mark options={color=red, thick}] (1.25, 1);})& 12.184 & 14.614 & 12.846\\

\cline{2-5}
&\textcircled{\tiny{10}} 4D-OS1024 (\tikz{\draw [dashed, color = red] (1, 1) -- (1.5, 1); \draw plot[mark=*, mark size=1.5,mark options={color=red, thick}] (1.25, 1);}) & 12.291 & 12.502 & 12.767\\

\cline{2-5}
&\textcircled{\tiny{11}} NL-4D-1024 (\tikz{\draw [dashed, color = brown] (1, 1) -- (1.5, 1); \draw plot[mark=x, mark size=2,mark options={color=brown, thick}] (1.25, 1);})& 12.417 & 12.586 & 13.085\\

\cline{2-5}
& Gaussian distribution &\multicolumn{2}{c|}{$\text{SNR}_{\text{req}}^{C}=11.761$}  
 & 12.143 \\
\hline
\hline
\end{tabular}
		}
  \label{tab:NLI_SNR}
  \vspace{-1.2em}
\end{table}

To evaluate the performance in the nonlinear optical fiber channel, we estimate the optimum effective SNR at a given distance  for each channel using the 4D  NLI model proposed in \cite{liang2023}. The 4D NLI model with considering modulation-dependent interference enables a quick computation of the NLI power $\sigma^2_{\text{NLI}}$ as a function of the input constellation for any  dual-polarization 4D format.

In Fig.~\ref{fig:NLI_results} (a) and (b), the optimal effective SNR $\text{SNR}_{\text{eff}}$ of each channel are shown for modulation formats with an SE of 6~bit/4D-sym and 8~bit/4D-sym, which listed in Table.~\ref{tab:NLI_SNR}, over multi-span  and single-span system, respectively. In order to compare the required and effective SNR differences    between the selected MI or GMI-optimal  modulation formats  in Fig.~\ref{fig:CapacityGap},  the $\text{SNR}_{\text{req}}^{\text{MI}}$, $\text{SNR}_{\text{req}}^{\text{GMI}}$ for AWGN channel and the $\text{SNR}_{\text{eff}}$ of center optical  channel are also shown in Table.~\ref{tab:NLI_SNR}. 
In addition,  we can also observe that the 4D-64PRS and NL-4D-1024 exhibit excellent nonlinear tolerance.
Despite the 2D-GS and 4D-GS designed for the AWGN channel have remarkable required SNR gains, they result in the degradation of the effective SNR due to the impact of nonlinear impairments caused during the propagation. 

\begin{figure}[!tb]
  \vspace{-1em}
    \centering
    {\includegraphics{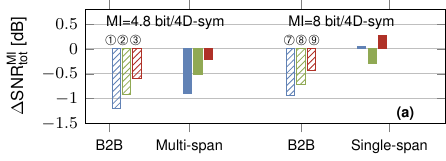}}
    {\includegraphics{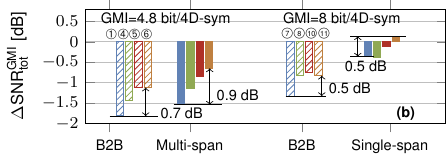}}
    \caption{The relative shaping gap or gain  of  the considered constellations in Table~\ref{tab:NLI_SNR} (marks as \textcircled{\tiny{1}} -- \textcircled{\tiny{11}}) with respect to Gaussian   distribution.
    }
    \label{fig:relative_gains}
    \vspace{-0.3em}
\end{figure}
To further quantify the  gains offered by 2D and 4D shaped  constellations in nonlinear optical fiber channel, 
the relative shaping gains in terms of the SNR ($\Delta \text{SNR}^{\text{MI}}_{\text{tot}}$ and $\Delta \text{SNR}^{\text{GMI}}_{\text{tot}}$) for the considered 2D and 4D modulation formats with respect to Gaussian distribution for both multi-span and single span scenarios are shown in Fig.~\ref{fig:relative_gains}. The total SNR gap or gains considering both linear and nonlinear effects  are defined as $\Delta\text{SNR}^R_{\text{tot}}=-\Delta \text{SNR}^R_{\text{req}}+\Delta \text{SNR}_{\text{eff}}$, where $\Delta \text{SNR}_{\text{eff}}$    represent the modulation-dependent effective SNR ($\text{SNR}_{\text{eff}}$) difference between a given constellation and Gaussian constellation itself in Table~\ref{tab:NLI_SNR}, resp.

 {In Fig.~\ref{fig:relative_gains}, 
for the back-to-back (B2B) case (pure AWGN channel),  the  NLI power equals zero, i.e., $\Delta \text{SNR}_{\text{eff}}=0$, thus $\Delta \text{SNR}^R_{\text{tot}}=-\Delta \text{SNR}^R_{\text{req}}$ (bar with lines). Note that all the formats reduce the gap to the theoretical AWGN-optimal Gaussian modulation in optical channel due to the largest NLI and lowest effective SNR of Gaussian modulation. 
In addition, 4D-GS formats  always provide positive  gains over uniform QAM, but 2D-GS may even have negative penalty (see the single-span case in Fig.~\ref{fig:relative_gains} (a)) with respect to QAM. 
In Fig.~\ref{fig:relative_gains} (b), the nonlinear-tolerant 4D  formats (brown bars) 
 can maintain or even provide  larger SNR gains  in the optical fiber channel than the gains in B2B cases. Thus, MD shaping is an efficient way to search for constellations which could tolerant the linear noise and nonlinear noise.
}

\vspace{-0.7em}
\section{Conclusion}
In this work, we evaluated the performance of various MD formats for both the AWGN channel and nonlinear optical fiber channel with two types of  transmission systems. The results suggest that more gains could be harvested with higher dimensional space for the AWGN channel and more nonlinear-tolerance could be found with four-dimensional space for the nonlinear optical channel. Therefore, higher dimensional space design may be the key to unlock the full potential of nonlinear constellation shaping.

\begin{spacing}{1}
{\small 
\linespread{1} \textbf{Acknowledgements}: 
This work is partially supported by the National Natural Science Foundation of China (62171175, 62001151), the Swedish Research Council (2021-03709), the  EuroTechPostdoc programme (754462) and the Netherlands Organisation for Scientific Research via the VIDI Grant  (15685). 
}
\end{spacing}


\printbibliography
\end{document}